\documentclass[epsfig,12pt]{article}
\usepackage{makeidx}
\usepackage{amsmath}
\usepackage{amsfonts}
\usepackage{amssymb}
\usepackage{graphicx}

\input epsf.sty
\newtheorem{theorem}{Theorem}
\newtheorem{acknowledgement}[theorem]{Acknowledgement}

\makeatletter \@addtoreset{equation}{section}
\renewcommand{\theequation}{\thesection.\arabic{equation}}
\input{tcilatex}
\begin{document}

\title{\vspace{-2cm}\rightline{\mbox{\small
{Lab/UFR-HEP0901/GNPHE/0901}}} \textbf{Tetrahedron in F-theory
Compactification }}
\author{El Hassan Saidi{\small \thanks{%
h-saidi@fsr.ac.ma}} \\
%EndAName
{\small \ Lab/UFR- Physique des Hautes Energies, Facult\'{e} des Sciences, }%
\\
{\small Rabat, Morocco,}}
\maketitle

\begin{abstract}
Complex tetrahedral surface $\mathcal{T}$ is a non planar projective surface
that is generated by four intersecting complex projective planes $CP^{2}$.
In this paper, we study the family $\left\{ \mathcal{T}_{m}\right\} $ of
blow ups of $\mathcal{T}$ and exhibit the link of these $\mathcal{T}_{m}$s
with the set of del Pezzo surfaces $dP_{n}$ obtained by blowing up n
isolated points in the $CP^{2}$. The $\mathcal{T}_{m}$s are toric surfaces
exhibiting a $U\left( 1\right) \times U\left( 1\right) $ symmetry that may
be used to engineer gauge symmetry enhancements in the Beasley-Heckman-Vafa
theory. The blown ups of the tetrahedron have toric graphs with faces, edges
and vertices where may localize respectively fields in adjoint
representations, chiral matter and Yukawa tri-fields couplings needed for
the engineering of F- theory GUT models building.\newline
\end{abstract}

%\tableofcontents

\section{Introduction}

With the advent of the Large Hadron Collider (LHC) at CERN, theoretical
studies around the Minimal Supersymmetric Standard model (\emph{MSSM}) and
Grand Unified Theories (GUT) have known intense activities. Among these
research activities, the studies of TeV- scale decoupled gravity scenarios
aiming the embedding of \emph{MSSM} and GUT models into superstrings and M-
theory \textrm{\cite{A1,A2,A3,A4}; }see also\textrm{\ \cite{A5,A6,A7,A8}}.
Recently Beasley-Heckman-Vafa made a proposal, to which we refer here below
as the BHV model, for embedding \emph{MSSM} and GUT into the 12D F-theory
compactified on Calabi-Yau four- folds \textrm{\cite{B1,B2,B3}}. In this
proposal, the visible supersymmetric gauge theory in 4D space time including
chiral matter and Yukawa couplings is given by an effective field model
following from the supersymmetric gauge theory on a seven brane wrapping 4-
cycles in the F-theory compactification down to \emph{4D} Minkowski space
time. In the engineering of the supersymmetric GUT models in the framework
of the BHV theory \textrm{\cite{B2,B3}}, see also \textrm{\cite{C1,C2}}, one
has to specify, amongst others, the geometric nature of the complex base
surface $S$ of the elliptically K3 fibered Calabi-Yau four -folds $X_{4}$:%
\begin{equation}
\begin{tabular}{lll}
$Y$ & $\rightarrow $ & $X_{4}$ \\ 
&  & $\downarrow \pi _{s}$ \\ 
&  & $S$%
\end{tabular}
\label{1}
\end{equation}%
In this relation Y is a complex two dimension fiber where live ADE
singularities giving rise to the rank $r$ gauge symmetry $G_{r}$ that we
observe in 4D space time and $S$ is a complex base surface whose cycle
homology captures important data on matter fields representations and their
tri- fields couplings.\ If the singular fiber Y of the local Calabi-Yau
four-folds (CY4) is fixed by the targeted 4D space time invariance $G_{r}$,
one may a priori imagine several kinds of compact complex surfaces $S$ as
its base manifold. The choice of $S$ depends on the effective 4D space time
physics; in particular the number of conserved supersymmetric charges and
chiral matter fields as well as their couplings. Generally speaking, the
simplest surfaces one may consider are likely those given by the so called
Hizerbruch surfaces $F_{n}=P^{1}\times _{n}P^{1}$ generated by fibration of
a complex projective line over a second projective line. Other examples of
surfaces are given by the complex projective plane $CP^{2}$ and its del
Pezzo $dP_{n}$ cousins; or in general non planar complex surfaces $\mathcal{D%
}$ embedded in higher dimensional complex Kahler manifolds. Typical examples
of adequate surfaces $S$ that have been explicitly studied in the BHV\ model
are given by the family of del Pezzo\emph{\ }surfaces $dP_{n}$\emph{\ }with $%
n=0,1,...,8$. These complex surfaces are obtained by preforming up to eight
blow ups at isolated points of the projective plane $CP^{2}=dP_{0}$ by
complex projective lines \textrm{\cite{DEL, DP, B1, DQ, DR}; }see also
section 2 for technical details.\newline
Motivated by the study of the geometric engineering of the F-theory GUT
models building \`{a} la BHV, we aim in this paper to contribute to this
matter by constructing a family of backgrounds for F-theory compactification
based on the tetrahedron geometry $\mathcal{T}$ and its blow ups. This study
sets up the basis for developing a class of F-theory \emph{GUT- like} models
building and uses the power of toric geometry of complex surfaces to
geometrically engineer chiral matter and the Yukawa couplings. Recall that
the tetrahedron $\mathcal{T}$ viewed as a toric surfaces with the following
toric fibration%
\begin{equation}
\begin{tabular}{lll}
$T^{2}$ & $\rightarrow $ & $\mathcal{T}$ \\ 
&  & $\downarrow \pi _{{\small \Delta }}$ \\ 
&  & $\Delta _{\mathcal{T}}$%
\end{tabular}%
\end{equation}%
has toric singularities generated by shrinking cycles of $T^{2}$ on the
edges of the tetrahedral base $\Delta _{\mathcal{T}}$ and at its vertices.
In our approach, the shrinking cycles of the above toric fibration are
interpreted in terms of gauge enhancement of bulk gauge symmetry $%
G_{r}\times U^{2}\left( 1\right) $. In going from a generic face of the
tetrahedron towards a vertex passing through a edge, the $G_{r}\times
U^{2}\left( 1\right) $ bulk gauge symmetry gets enhanced to $G_{r+1}\times
U\left( 1\right) $ on the edge and to $G_{r+2}$ at the vertex as shown on
the following table:

\begin{equation*}
\begin{tabular}{|lllllllll|}
\hline
&  &  &  &  &  &  &  &  \\ 
& {\small Tetrahedron} $\mathcal{T}$ & : & \ \ \ \ {\small faces} & \ \ \ \
\  & {\small edges} & \ \ \ \ \  & {\small vertices} &  \\ 
& {\small toric symmetry} & : & ${\small U}\left( 1\right) {\small \times U}%
\left( 1\right) $ &  & ${\small U}\left( 1\right) $ &  & \ \ {\small -} & 
\\ 
& {\small gauge enhancement} & : & $\ \ \ \ {\small G}_{r}\times {\small U}%
^{2}\left( 1\right) $ &  & ${\small G}_{r+1}\times {\small U}\left( 1\right) 
$ &  & ${\small G}_{r+2}$ &  \\ 
&  &  &  &  &  &  &  &  \\ \hline
\end{tabular}%
\end{equation*}

\ \ \newline
In the present paper, we focus our attention mainly on the study of the
typical family of base surfaces $S$ of eq(\ref{1}) involving the non planar
complex tetrahedral surface and its blow ups denoted here below as $\mathcal{%
T}_{{\small 0}}$ and $\mathcal{T}_{n}$ respectively. In the conclusion
section, we give comments on the engineering of GUT-like 4D $\mathcal{N}=1$
supersymmetric quiver gauge models based on $\mathcal{T}_{{\small 0}}$ and $%
\mathcal{T}_{n}$. A more involved and explicit study for the engineering of
F- theory GUT-like models along the line of the BHV approach; but now with $%
\mathcal{T}_{{\small 0}}$ and $\mathcal{T}_{n}$ as complex base geometries
in the local Calabi-Yau four-folds of eq(\ref{1}) will be reported in 
\textrm{\cite{DMS}}.\newline
The presentation of this paper is as follows: In section 2, we review
general aspects of del Pezzo surfaces $dP_{k}$; in particular their 2- cycle
homology classes and their links to the exceptional\textrm{\footnote{%
Here $E_{3}$, $E_{4}$, $E_{5}$ denote respectively $SU\left( 3\right) \times
SU\left( 2\right) $, $SU\left( 5\right) $ and $SO\left( 10\right) $ and $%
E_{6}$, $E_{7}$, $E_{8}$ are the usual exceptional Lie algebras in Cartan
classification.}} Lie algebras. This review on real 2- cycle homology of the
dP$_{k}$s is important to shed more light for the study and the building of
the blow ups of the tetrahedron. In section 3, we introduce the complex
tetrahedral surface $\mathcal{T}_{{\small 0}}$; first as a complexification
of the usual real tetrahedron (hollow triangular pyramid); that is as a non
planar complex surface given by the intersection of four projective planes $%
CP^{2}$. Second as a complex codimension one divisor ( "a toric boundary")
of the complex three dimension projective space $CP^{3}$. We take also this
opportuinity to recall useful results on $CP^{3}$ thought of as a toric
manifold and its Chern classes $c_{k}\left( {\small TCP}^{{\small 3}}\right) 
$. These tools are used in section 4 to study the blow ups of the
tetrahedron; in particular the toric blow ups of its vertices by projective
planes and the blow up of its edges by the del Pezzo surface dP$_{1}$. In
section 5, we give a conclusion and make comments on supersymmetric GUT-like
quiver gauge theories embedded in F-theory compactification on local
Calabi-Yau four-folds.

\section{Del Pezzo surfaces $dP_{k}$}

We first consider the 2- cycle homology of the del Pezzo surfaces. Then we
give the links between these surfaces and the roots system of the
"exceptional" Lie algebras.

\subsection{Homology of $dP_{k}$}

\begin{figure}[tbph]
\begin{center}
\hspace{0cm} \includegraphics[width=14cm]{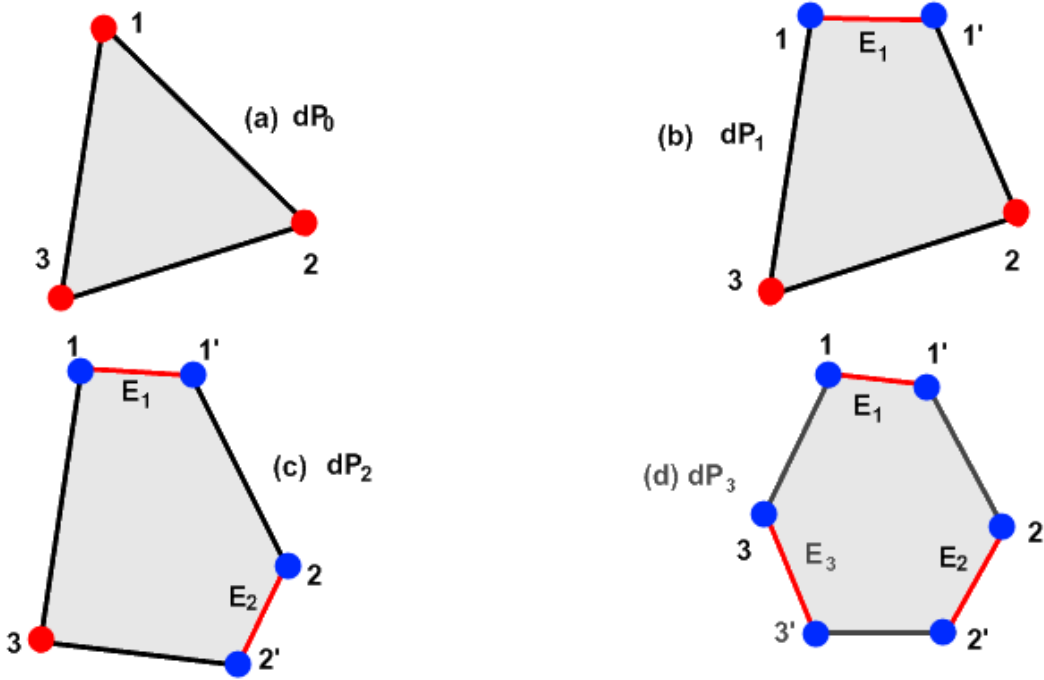}
\end{center}
\par
\vspace{-0.5 cm}
\caption{{\protect\small Toric graphs for dP}$_{0}${\protect\small \ , dP}$%
_{1}${\protect\small \ , dP}$_{2}$ {\protect\small and dP}$_{3}$%
{\protect\small . The surface dP}$_{1}$ is obtained by blowing up the vertex
1. The other are recovered by blowing up the vertices 2 and 3.}
\label{del}
\end{figure}

The $dP_{k}$ del Pezzo surfaces with $k\leq 8$ are defined as blow ups of
the complex projective space $CP^{2}$ at $k$ points. Taking into account the
overall size $r_{0}$ of the compact $CP^{2}$, a surface $dP_{k}$ has then
real $\left( k+1\right) $ dimensional Kahler moduli $\left\langle
r_{0},r_{1},\ldots ,r_{k}\right\rangle $ corresponding to the volume of each
blown up cycle. The $dP_{k}$s possess as well a moduli space of complex
structures with complex dimension $\left( 2k-8\right) $ where the eight
gauge fixed parameters are associated with the $GL\left( 3\right) $ symmetry
of $CP^{2}$. As such, only surfaces with $5\leq k\leq 8$ admit a moduli
space of complex structures.\newline
The real 2-cycle homology group $\mathbb{H}_{2}\left( dP_{k},Z\right) $ is $%
\left( k+1\right) $ dimensional and is generated by $\left\{
H,E_{1},...,E_{k}\right\} $ where $H$ denotes the hyperplane class inherited
from $CP^{2}$ and the $E_{i}$ denote the exceptional divisors associated
with the blow ups. These generators have the intersection pairing 
\begin{equation}
\begin{tabular}{llllll}
$H.H=1$ & , & $H.E_{i}=0$ & , & $E_{i}.E_{j}=-\delta _{ij}\quad ,\quad
i,j=1,...,k$ & ,%
\end{tabular}
\label{he}
\end{equation}%
so that the signature $\eta $ of the $\mathbb{H}_{2}\left( dP_{k},Z\right) $
group is given by ${\small diag}\left( +-...-\right) $.\newline
The first three blow ups giving $dP_{1},$ $dP_{2}$ and $dP_{3}$ complex
surfaces are of toric types while the remaining five others namely $%
dP_{4},...,$ $dP_{8}$ are non toric. These projective surfaces have the
typical toric fibration 
\begin{equation}
\begin{tabular}{lll}
$T^{2}$ & $\rightarrow $ & $dP_{k}$ \\ 
&  & $\downarrow $ \\ 
&  & $\emph{B}_{2,k}$%
\end{tabular}%
,\qquad k=1,2,3,
\end{equation}%
with real real two dimension base $\emph{B}_{2,k}$ nicely represented by
toric diagrams $\Delta _{2,k}$ encoding the toric data on the shrinking
cycles in the toric fibration%
\begin{equation*}
\begin{tabular}{ll|l|ll|ll|ll}
{\small surface S} &  & ${\small dP}_{0}{\small =CP}^{2}$ &  & ${\small dP}%
_{1}$ &  & ${\small dP}_{2}$ &  & ${\small dP}_{3}$ \\ \hline
{\small blow ups} &  & ${\small k=0}$ &  & ${\small k=1}$ &  & ${\small k=2}$
&  & ${\small k=3}$ \\ 
{\small toric graph }$\Delta _{2,k}$ &  & {\small triangle} &  & {\small %
quadrilateral} &  & {\small pentagon} &  & {\small hexagon} \\ 
{\small generators} &  & ${\small H}$ &  & ${\small H}$ ${\small ,}$ $%
{\small E}_{1}$ &  & ${\small H}$ ${\small ,}$ ${\small E}_{1}$ ${\small ,}$ 
${\small E}_{2}$ &  & ${\small H}$ ${\small ,}$ ${\small E}_{1}$ ${\small ,}$
${\small E}_{2}$ ${\small ,}$ ${\small E}_{2}$ \\ \hline
\end{tabular}%
\end{equation*}

\ \ \newline
The toric graphs of the projective plane ${\small CP}^{2}$ and its toric
blown ups namely $dP_{1}$, $dP_{2}$ and $dP_{3}$ are depicted in the figure (%
\ref{del}). The surfaces $dP_{k}$ with $4\leq k\leq 8$ have no toric graph
representation.

In terms of the basic hyperline $H$ and the exceptional curves $E_{i}$,
generic classes $\left[ \Sigma _{a}\right] $ of complex holomorphic curves
in the del Pezzos $dP_{k}$ are given by the following integral linear
combinations, 
\begin{equation}
\Sigma _{a}=n_{a}H-\sum_{i=1}^{k}m_{ai}E_{i},  \label{ca}
\end{equation}%
with $n_{a}$ and $m_{a}$ are integers. The self- intersection numbers $%
\Sigma _{a}^{2}\equiv \Sigma _{a}\cdot \Sigma _{a}$ following from eqs(\ref%
{ca}) and (\ref{he}) are then given by 
\begin{equation}
\Sigma _{a}^{2}=n_{a}^{2}-\sum_{i=1}^{k}m_{ai}^{2}.  \label{c}
\end{equation}%
The canonical class $\Omega _{k}$ of the projective $dP_{k}$ surface, which
is given by \emph{minus} the first Chern class $c_{1}\left( dP_{k}\right) $
of the tangent bundle of the surface $dP_{k}$, reads as, 
\begin{equation}
\Omega _{k}=-\left( 3H-\sum_{i=1}^{k}E_{i}\right) ,
\end{equation}%
and has a self intersection number $\Omega _{k}^{2}=9-k$ whose positivity
requires $k<9$. Obviously $k=0$ corresponds just to the case where there is
no blow up.\ The degree $d_{\Sigma }$ of a generic complex curve class $%
\Sigma =nH-\sum_{i=1}^{k}m_{i}E_{i}$ in $dP_{k}$ is given by the
intersection number between the class $\Sigma $ with the anticanonical class 
$\left( -\Omega _{k}\right) $, 
\begin{equation}
d_{\Sigma }=-\text{ }\left( \Sigma \cdot \Omega _{k}\right)
=3n-\sum_{i=1}^{k}m_{i}.
\end{equation}%
Positivity of this integer $d_{\Sigma }$ puts a constraint equation on the
allowed values of the $n$ and \ $m_{i}$ integers which should be like,%
\begin{equation}
\sum_{i=1}^{k}m_{i}\leq 3n.  \label{ccs}
\end{equation}%
Notice that there is a remarkable relation between the self intersection
number $\Sigma ^{2}$ (\ref{c}) of the classes of holomorphic curves and
their degrees $d_{\Sigma }$. This relation, which is known as the \emph{%
adjunction formula }\textrm{\cite{AD, DEL}}, is given by $\Sigma
^{2}=2g-2+d_{\Sigma }$, and allows to define the genus $g$ of the curve
class $\Sigma $ as 
\begin{equation}
g=1+\frac{n\left( n-3\right) }{2}-\sum_{i=1}^{k}\frac{m_{i}\left(
m_{i}-1\right) }{2}.
\end{equation}%
For instance, taking $\Sigma =3H$; that is $n=3$ and $m_{i}=0$, then the
genus $g_{3H}$ of this curve is equal to $1$ and so the curve $3H$ is in the
same class of the real 2- torus. In general, fixing the genus $g$ to a given
positive integer puts then a second constraint equation on $n$ and $m_{i}$
integers; the first constraint is as in (\ref{ccs}). For the example of
rational curves with $g=0$, we have $\Sigma ^{2}=d_{\Sigma }-2$ giving a
relation between the degree $d_{\Sigma }$ of the curve $\Sigma $ and its
self intersection. For $d_{\Sigma }=0$, we have a rational curve with self
intersection $\Sigma ^{2}=-2$ while for $d_{\Sigma }=1$ we have a self
intersection $\Sigma ^{2}=-1$. To get the general expression of genus $g=0$
curves, one has to solve the constraint equation $\sum_{i=1}^{k}m_{i}\left(
m_{i}-1\right) =2+n\left( n-3\right) $ by taking into account the condition (%
\ref{ccs}). For $k=1$, this relation reduces to $m\left( m-1\right)
=2+n\left( n-3\right) $, its leading solutions $n=1,$ $m=0$ and $n=0,$ $m=-1$
give just the classes $H$ and $E$ respectively with degrees $d_{H}=3$ and $%
d_{E}=1$. Typical solutions for this constraint equation are given by the
generic class $\Sigma _{n,n-1}=nH-\left( n-1\right) E$ which is more
convenient to rewrite it as follows $\Sigma _{n,n-1}=H+\left( n-1\right)
\left( H-E\right) $.

\subsection{Link to roots of Lie algebras}

Del Pezzo surfaces $dP_{k}$ have also a remarkable link with the exceptional
Lie algebras. Decomposing the $\mathbb{H}_{2}$ homology group like, 
\begin{equation}
\begin{tabular}{llll}
$\mathbb{H}_{2}\left( dP_{k},Z\right) _{k\geq 3}$ & $=$ & $\left\langle
\Omega _{k}\right\rangle \oplus \mathcal{L}_{k}$ & ,%
\end{tabular}
\label{h}
\end{equation}%
with%
\begin{equation}
\begin{tabular}{llll}
$\Omega _{k}$ & $=$ & $-3H+E_{i}+\cdots +E_{k}$ & , \\ 
$\mathcal{L}_{k}$ & $=$ & $\left\langle \Omega _{k}\right\rangle ^{\bot }$ & 
,%
\end{tabular}%
\end{equation}%
where the sublatice $\mathcal{L}_{k}=\left\langle \alpha _{1},...,\alpha
_{k}\right\rangle $, which is orthogonal to\ $\Omega _{k}$, is identified
with the root space of the corresponding Lie algebra $E_{k}$. The generators 
$\alpha _{i}$ of the lattice $\mathcal{L}_{k}$ are: 
\begin{equation}
\begin{tabular}{ll}
$\alpha _{{\small 1}}=E_{{\small 1}}-E_{{\small 2}}$ & , \\ 
$\ \ \ \ \ \ \ \ \ \ \ \ \ \vdots $ &  \\ 
$\alpha _{{\small k-1}}=E_{{\small k-1}}-E_{{\small k}}$ & , \\ 
$\alpha _{{\small k}}=H-E_{1}-E_{2}-E_{3}$ & ,%
\end{tabular}
\label{ha}
\end{equation}%
with pairing product $\alpha _{i}.\alpha _{j}$ equal to minus the Cartan
matrix $C_{ij}\left( E_{k}\right) $ of the Lie algebra E$_{k}$. For the
particular case of $dP_{2}$, the corresponding Lie algebra is $su\left(
2\right) $. The mapping between the exceptional curves and the roots of the
exceptional Lie algebras is given in the following table%
\begin{equation}
\text{%
\begin{tabular}{ll|ll}
dP$_{k}$ surfaces & exceptional curves & Lie algebras & simple roots \\ 
\hline\hline
$dP_{1}$ & \multicolumn{1}{|l|}{${\small E}_{1}$} & - & \multicolumn{1}{|l}{-
} \\ 
$dP_{2}$ & \multicolumn{1}{|l|}{${\small E}_{1},${\small \ }${\small E}_{2}$}
& $su\left( 2\right) $ & \multicolumn{1}{|l}{${\small \alpha }_{{\small 1}}$}
\\ 
$dP_{3}$ & \multicolumn{1}{|l|}{${\small E}_{1},${\small \ }${\small E}_{2},$%
{\small \ }${\small E}_{3}$} & $su\left( 3\right) \times su\left( 2\right) $
& \multicolumn{1}{|l}{${\small \alpha }_{{\small 1}},${\small \ }${\small %
\alpha }_{{\small 2}},${\small \ }${\small \alpha }_{{\small 3}}$} \\ 
$dP_{4}$ & \multicolumn{1}{|l|}{${\small E}_{1},${\small \ }${\small E}_{2},$%
{\small \ }${\small E}_{3},{\small E}_{4}$} & $su\left( 5\right) $ & 
\multicolumn{1}{|l}{${\small \alpha }_{{\small 1}},${\small \ }${\small %
\alpha }_{{\small 2}},${\small \ }${\small \alpha }_{{\small 3}},$ ${\small %
\alpha }_{{\small 4}}$} \\ 
$dP_{5}$ & \multicolumn{1}{|l|}{${\small E}_{1},${\small \ }${\small E}_{2},$%
{\small \ }${\small E}_{3},$ ${\small E}_{4},${\small \ }${\small E}_{5}$} & 
$so\left( 10\right) $ & \multicolumn{1}{|l}{${\small \alpha }_{{\small 1}},$%
{\small \ }${\small \alpha }_{{\small 2}},${\small \ }${\small \alpha }_{%
{\small 3}},${\small \ }${\small \alpha }_{{\small 4}},{\small \alpha }_{%
{\small 5}}$} \\ 
$dP_{6},dP_{7},dP_{8}$ & \multicolumn{1}{|l|}{${\small E}_{1},{\small \ E}%
_{2},{\small \ ...},$ ${\small E}_{k}$} & $E_{6},E_{7},E_{8}$ & 
\multicolumn{1}{|l}{${\small \alpha }_{{\small 1}},${\small \ ..., }${\small %
\alpha }_{{\small k}},$ ${\small k=6,7,8}$} \\ \hline
\end{tabular}%
}  \label{exp}
\end{equation}

\ \ \newline
Notice that one can also use eqs(\ref{h},\ref{ha}) to express the generators 
$H$ and $\left\langle E_{i}\right\rangle _{1\leq i\leq k}$ in terms of the
anticanonical class $\Omega _{k}$ and the roots of the exceptional Lie
algebra; for details see \textrm{\cite{DMS}}.

\section{Tetrahedral surface}

The complex tetrahedral surface $\mathcal{T}_{{\small 0}}$ has much to do
with the usual real triangular hollow\textrm{\footnote{%
One should distinguish two kinds of triangular pyramids: filled and empty.
We are interested in the second one denoted as $\Delta _{\mathcal{T}_{%
{\small 0}}}$. The real triangular pyramid with filled bulk is denoted by $%
\Delta _{{\small CP}^{{\small 3}}}$ ; it is the toric graph of $CP^{3}$. We
also have the relation $\Delta _{\mathcal{T}_{{\small 0}}}=\partial \left(
\Delta _{{\small CP}^{{\small 3}}}\right) $.}} pyramid which we denote as $%
\Delta _{\mathcal{T}_{{\small 0}}}$. In this section, we want to exhibit
explicitly this link; but also its relation to the complex three dimension
projective space $CP^{3}$. To that purpose, we first describe the relation
between the complex tetrahedral surface $\mathcal{T}_{{\small 0}}$ and the
complex projective plane $CP^{2}$. Then we examine its relation with the
complex three dimension space $CP^{3}$. Because of the link between $%
\mathcal{T}_{{\small 0}}$ and $CP^{3}$, we take this occasion to give useful
results on the homology of $CP^{3}$ which we use in section 4 to study the
blowing up of the tetrahedron.

\subsection{Link between $\mathcal{T}_{{\protect\small 0}}$ and $CP^{2}$}

\begin{figure}[tbph]
\begin{center}
\hspace{0cm} \includegraphics[width=5cm]{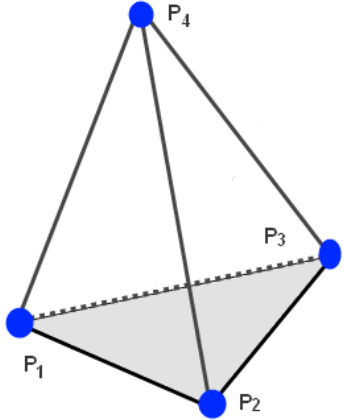}
\end{center}
\par
\vspace{-0.5 cm}
\caption{This figure represents the toric graph $\Delta _{\mathcal{T}_{%
{\protect\small 0}}}$ of the complex tetrahedral surface $\mathcal{T}_{%
{\protect\small 0}}$. This toric surface is a candidate for a base surface
of local CY4s in the BHV theory. On the faces of $\Delta _{\mathcal{T}_{%
{\protect\small 0}}}$ live fields in adjoint representation of $G_{r}\times
U^{2}\left( 1\right) $, while on the edges lives bi- fundamentals and at
vertices it lives tri- fields Yukawa couplings}
\label{BV}
\end{figure}

Roughly, the complex tetrahedral surface $\mathcal{T}_{{\small 0}}$ extends
the complex projective plane $CP^{2}$; it is a non planar projective surface
that involve several projective planes $\left\{ CP_{a}^{2}\right\} $ and
whose basic properties may be read from the real tetrahedron $\Delta _{%
\mathcal{T}_{{\small 0}}}$. The latter is given by the four external faces
of the triangular hollow pyramid $\Delta _{\mathcal{T}_{{\small 0}}}$ whose
graph is depicted in (\ref{BV}).

\ \ \newline
Form the figures (\ref{del}) and (\ref{BV}) as well as the relation between
triangles\textrm{\footnote{%
In toric geometry, projective lines $CP^{1}$ are presented by segments $%
\left[ AB\right] $, projective planes $CP^{2}$ by triangles $\left[ ABC%
\right] $ and in general $CP^{n}$ spaces by n-simplex $\left[ A_{1}...A_{n+1}%
\right] $.}} and projective planes, we immediately learn that there is a
strong link between the usual tetrahedron $\Delta _{\mathcal{T}_{{\small 0}%
}} $ \ and the complex tetrahedral surface $\mathcal{T}_{{\small 0}}$. This
non planar surface is then built in terms of four intersecting compact
projective planes $CP_{1}^{2}$, $CP_{2}^{2}$, $CP_{3}^{2}$ and $CP_{4}^{2}$
which are in one to one correspondence with the four faces of $\Delta _{%
\mathcal{T}_{{\small 0}}}$. The intersection of any two projective planes;
say $CP_{a}^{2}$ and $CP_{b}^{2}$, is a complex projective line $\Sigma
_{\left( ab\right) }\sim CP^{1}$ and are\ associated with the edges of $%
\Delta _{\mathcal{T}_{{\small 0}}}$;%
\begin{equation}
\begin{tabular}{lll}
$\Sigma _{\left( ab\right) }$ & $=CP_{a}^{2}\cap CP_{b}^{2}$ & , \\ 
$\Sigma _{\left( ab\right) }$ & $\simeq CP^{1}$ & ,%
\end{tabular}
\label{si}
\end{equation}%
with $b>a=1,...,4$. Moreover we learn also that any triplet of three
projective planes; say $CP_{a}^{2}$, $CP_{b}^{2}$ and $CP_{c}^{2}$, meet at
one of the four vertices of the tetrahedron, i.e:%
\begin{equation}
\begin{tabular}{llll}
$P_{\left( abc\right) }$ & $=$ & $CP_{a}^{2}\cap CP_{b}^{2}\cap CP_{c}^{2}$
& .%
\end{tabular}%
\end{equation}%
Up on using eq(\ref{si}) may be also written as%
\begin{equation}
\begin{tabular}{llll}
$P_{\left( abc\right) }$ & $=$ & $\Sigma _{\left( ab\right) }\cap CP_{c}^{2}$
& $,$ \\ 
& $=$ & $\Sigma _{\left( bc\right) }\cap CP_{a}^{2}$ & , \\ 
& $=$ & $\Sigma _{\left( ac\right) }\cap CP_{b}^{2}$ & ,%
\end{tabular}%
\end{equation}%
with $c>b>a=1,...,4$. These vertices may be as well defined as the
intersection of edges $\Sigma _{\left( ab\right) }$ and $\Sigma _{\left(
bc\right) }$ or equivalently $\Sigma _{\left( bc\right) }$ and $\Sigma
_{\left( ac\right) }$.\newline
Notice that the exact link between $\mathcal{T}_{{\small 0}}$ and $\Delta _{%
\mathcal{T}_{{\small 0}}}$ is given by toric geometry which allows to define
the complex tetrahedral surface $\mathcal{T}_{{\small 0}}$ in terms of the
following toric fibration,%
\begin{equation}
\begin{tabular}{lll}
$T_{\mathcal{T}_{{\small 0}}}^{2}$ & $\rightarrow $ & $\mathcal{T}_{{\small 0%
}}$ \\ 
&  & $\downarrow \pi $ \\ 
&  & $B_{\mathcal{T}_{{\small 0}}}$%
\end{tabular}%
\end{equation}%
where the fiber $T^{2}$ stands for the 2- torus $S^{1}\times S^{1}$ and $B_{%
\mathcal{T}_{{\small 0}}}$ for a real two dimensional base. The polytope $%
\Delta _{\mathcal{T}_{{\small 0}}}$ is precisely the toric graph of the real
base $B_{\mathcal{T}_{{\small 0}}}$. This toric graph encodes the toric data
of the toric symmetries of the complex tetrahedral surface viewed as a
complex two dimension toric manifold. As these toric data are intimately
related to the toric representation of CP$^{3}$; we give these details in
the next section.

\subsection{Relation between $\mathcal{T}_{{\protect\small 0}}$ and $CP^{3}$}

Along with its connection with $CP^{2}$, the complex tetrahedral surface $%
\mathcal{T}_{{\small 0}}$ has as well a strong link with the complex three
dimension projective space $CP^{3}$. The projective planes $CP_{1}^{2}$, $%
CP_{2}^{2}$, $CP_{3}^{2}$ and $CP_{4}^{2}$ encountered in the previous
subsection are precisely the four basic divisors $D_{1}$, $\mathcal{D}_{2}$, 
$\mathcal{D}_{3}$ and $\mathcal{D}_{4}$ of $CP^{3}$. In terms of the
holomorphic coordinates $\left\{ x_{a}\right\} $ of the complex four
dimension space $C^{4}$ where live $CP^{3}$, we can define these basic
divisors $\mathcal{D}_{a}$ by the following hypersurfaces, 
\begin{equation}
\mathcal{D}_{a}=\left\{ 
\begin{array}{c}
\left( x_{1},x_{2},x_{3},x_{4}\right) \equiv \left( \lambda x_{1},\lambda
x_{2},\lambda x_{3},\lambda x_{4}\right) \text{ } \\ 
\left( x_{1},x_{2},x_{3},x_{4}\right) \neq \left( 0,0,0,0\right) \text{ and }%
x_{a}=0%
\end{array}%
\right\}  \label{d}
\end{equation}%
with $a=1,2,3,4$ and where $\lambda $ is a non zero complex number; the
parameter of the $C^{\ast }$ action. In this set up, the complex tetrahedral
surface may be defined as the complex codimension one hypersurface%
\begin{equation}
\begin{tabular}{llll}
$\mathcal{T}_{{\small 0}}$ & $=$ & $\dbigcup_{a=1}^{4}\mathcal{D}_{a},$ & ,%
\end{tabular}
\label{dd}
\end{equation}%
together with the following bi- and tri- intersections%
\begin{equation}
\begin{tabular}{llll}
$\Sigma _{\left( ab\right) }$ & $=$ & $\mathcal{D}_{a}\cap \mathcal{D}%
_{b}\quad ,\quad a<b$ & , \\ 
$P_{\left( abc\right) }$ & $=$ & $\mathcal{D}_{a}\cap \mathcal{D}_{b}\cap 
\mathcal{D}_{c}\quad ,\quad a<b<c$ & .%
\end{tabular}%
\end{equation}%
Notice in passing that in the effective 4D space time physics of branes
wrapping cycles in type II strings on Calabi Yau threefolds and F-theory on
CY4-folds, these cycles intersections give rise to branes intersections
which have a nice interpretation in terms of chiral matter in
bi-fundamentals and tri-fields couplings. \newline
In the toric geometry language, the complex tetrahedral surface $\mathcal{T}%
_{{\small 0}}$ is in some sens\textrm{\footnote{%
What we mean by the toric boundary of a complex n dimension manifold $M_{n}$
is the codimension one toric submanifold $M_{n-1}=\partial \left(
M_{n}\right) _{toric}$ associated with the shrinking of then n-torus fiber $%
T^{n}$ of $M_{n}$ down to $T^{n-1}$.}} the "toric boundary" of $CP^{3}$.
Recall that $CP^{3}$ is a toric manifold with the toric fibration%
\begin{equation}
\begin{tabular}{lll}
$T^{3}$ & $\rightarrow $ & $CP^{3}$ \\ 
&  & $\downarrow \pi $ \\ 
&  & $B_{3}$%
\end{tabular}
\label{fbb}
\end{equation}%
where the real three dimension base $B_{3}$ has as a toric polytope given by
the 3- simplex $\Delta _{CP^{3}}$. This 3- simplex is just the triangular
pyramid with filled bulk and is related to $\Delta _{\mathcal{T}_{{\small 0}%
}}$ as follows, 
\begin{equation}
\Delta _{\mathcal{T}_{{\small 0}}}=\partial \left( \Delta _{{\small CP}^{%
{\small 3}}}\right) .
\end{equation}%
As such the complex tetrahedral surface $\mathcal{T}_{{\small 0}}$ inherits
specific features of the toric data of the complex projective space $CP^{3}$%
. These toric data, which are encoded on the faces, the edges and the
vertices of the polytope $\Delta _{CP^{3}}$, are generated by shrinking
cycles of the $T^{3}$ fiber of eq(\ref{fbb}). In the next section we will
use these data to study the toric blown ups of $\mathcal{T}_{{\small 0}}$;
but before that let us complete this discussion by recalling useful results
on the Chern classes for $CP^{3}$. These classes may be read from the total
Chern class given by the following sum%
\begin{equation}
c_{tot}\left( X\right) =1+c_{1}\left( X\right) +c_{2}\left( X\right)
+c_{3}\left( X\right) ,  \label{38}
\end{equation}%
where $X$ stands for complex three dimension manifold and where the $%
c_{k}\left( X\right) $ refer to $c_{k}\left( TX\right) $; i.e the k-th Chern
class of the tangent bundle $TX$. \newline
For the case $X=CP^{3}$, the Chern classes $c_{k}\left( X\right) $ are
generated by a single two dimensional class $\omega $ reads as follows%
\begin{equation}
\begin{tabular}{llll}
$c_{tot}\left( X\right) $ & $=$ & $\left( 1+\omega \right) ^{4}=1+4\omega
+6\omega ^{2}+4\omega ^{3}$ & ,%
\end{tabular}
\label{39}
\end{equation}%
together with the normalization 
\begin{equation}
\int_{CP^{3}}\omega ^{3}=1  \label{93}
\end{equation}%
and the nilpotent relation $\omega ^{4}=0$. From the relations (\ref{38})
and (\ref{39}), we can read directly the expression of the first $%
c_{1}\left( X\right) $, the second $c_{2}\left( X\right) $ and the third $%
c_{3}\left( X\right) $ Chern classes,%
\begin{equation}
\begin{tabular}{llll}
$c_{1}\left( X\right) $ & $=$ & $4\omega $ & , \\ 
$c_{2}\left( X\right) $ & $=$ & $6\omega ^{2}$ & , \\ 
$c_{3}\left( X\right) $ & $=$ & $4\omega ^{3}$ & ,%
\end{tabular}%
\end{equation}%
as well as the Euler characteristic 
\begin{equation}
\chi \left( X\right) =\int_{CP^{3}}c_{3}\left( X\right) =4
\end{equation}%
in agreement with the Gauss Bonnet theorem for $CP^{3}$. Notice that
expressing the normalization condition (\ref{93}) like 
\begin{equation}
\int_{CP^{3}}\omega \wedge \omega ^{2}=1,
\end{equation}%
one learns amongst others that that real 2- forms and real 4- forms are dual
in $CP^{3}$. The same duality is valid for real 2- cycles $\Sigma $ and
codimension 2 real 4-cycles $D$ that satisfy the following pairings:%
\begin{equation}
\begin{tabular}{llll}
$\left\langle \Sigma ,D\right\rangle _{CP^{3}}$ & $=$ & $1$ & , \\ 
$\int_{D}\omega ^{2}$ & $=$ & $1$ & , \\ 
$\int_{\Sigma }\omega $ & $=$ & $1$ & ,%
\end{tabular}%
\end{equation}%
Notice moreover that the 2-form $\omega $ is the curvature of a line bundle $%
\mathcal{L}^{\ast }$ whose complex conjugate $\mathcal{L}$ is precisely the
generating line bundle over $CP^{3}$ with total Chern class, 
\begin{equation}
c_{tot}\left( \mathcal{L}\right) =1-\omega .  \label{40}
\end{equation}%
A remarkable line bundle over $CP^{3}$ is given by the maximum exterior
power of the cotangent bundle $T^{\ast }X$ with $X=CP^{3}$. This is the
canonical line bundle 
\begin{equation}
\mathcal{K}=\left( T^{\ast }X\right) \wedge \left( T^{\ast }X\right) \wedge
\left( T^{\ast }X\right)
\end{equation}
whose Chern class given by $c_{tot}\left( \mathcal{K}\right) =1-4\omega $
with $c_{1}\left( \mathcal{K}\right) =-4\omega $. From these relations, we
learn that $\mathcal{K}$ is the fourth power of the generating line bundle $%
\mathcal{L}$, 
\begin{equation}
\mathcal{K}=\mathcal{L}^{4}
\end{equation}%
We learn as well that $c_{1}\left( \mathcal{K}\right) $ is nothing but $%
c_{1}\left( T^{\ast }X\right) =-c_{1}\left( TX\right) $.

\section{Blown up geometries}

First notice that the blow ups of the tetrahedral surface may be classified
in two types: toric blow ups and non toric ones. In this section, we will
mainly focus on the toric blow ups which can be\ engineered directly from
the toric graph $\Delta _{\mathcal{T}_{{\small 0}}}$ given by the figure (%
\ref{BV}). \newline
Moreover, within the class of toric blow ups, we also distinguish two
subsets of toric blow ups of $\mathcal{T}_{{\small 0}}$ depending on the
dimension of the shrinking cycles:\newline
(\textbf{1}) blow ups of the four vertices of $\Delta _{\mathcal{T}_{{\small %
0}}}$ in terms of projective planes $CP^{2}$. These are the analogs of the
blow ups we encounter in building del Pezzo surfaces form the projective
plane. They are associated with singularities at isolated points.\newline
(\textbf{2}) blow ups the edges of $\Delta _{\mathcal{T}_{{\small 0}}}$ by
using projective lines $CP^{1}$. This kind of blow ups has no analog in the
blowing up of $CP^{2}$. The blown ups surfaces will be denoted as $\mathcal{T%
}_{k}^{\prime }$.\newline
Recall that at the four vertices of the tetrahedron $\Delta _{{\small CP}^{%
{\small 3}}}$, a 3-torus $T^{3}$ shrinks to zero%
\begin{equation}
\begin{tabular}{llllll}
$vertex$ & : & $T^{3}$ & $\rightarrow $ & $0$ & ,%
\end{tabular}%
\end{equation}%
while on its six the edges we have shrinking 2-tori.%
\begin{equation}
\begin{tabular}{llllll}
$edge$ & : & $T^{2}$ & $\rightarrow $ & $0$ & .%
\end{tabular}%
\end{equation}%
Below, we study these two kinds of blow ups by first considering blowing ups
by projective planes by essentially mimicking the building of del Pezzo
surfaces in terms of the blow ups of $CP^{2}$ considered in section 2.

\subsection{Blow ups of points by CP$^{2}$s}

\begin{figure}[tbph]
\begin{center}
\hspace{0cm} \includegraphics[width=4cm]{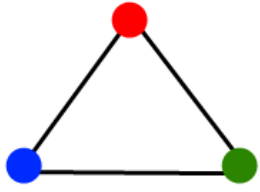}
\end{center}
\par
\vspace{-0.5 cm}
\caption{The Dynkin diagram of the affine A$_{{\protect\small 2}}$\
singularity. Viewed as the intersecting of three divisors $D_{i}\sim
T_{i}^{2}\times \Delta _{i}$, the singularity at the vertex $%
P_{abc}=D_{a}\cap D_{b}\cap D_{c}$ involves the simultaneous shrinking of
the three $T_{i}^{2}$s to zero. Each node is associated with the shrinking
of one of the $T_{i}^{2}$s interpreted as an ordinary A$_{{\protect\small 1}%
} $\ singularity.{\protect\small \ }}
\label{A2}
\end{figure}

To start notice that by thinking about the complex tetrahedral surface $%
\mathcal{T}_{{\small 0}}$ $\sim $ $T^{2}\times \Delta _{\mathcal{T}_{{\small %
0}}}$ as a toric submanifold of $CP^{3}$ $\sim $ $T^{3}\times \Delta _{%
{\small CP}^{{\small 3}}}$, that is roughly as its toric boundary; see also
footnote (2),%
\begin{equation}
\begin{tabular}{llll}
$\Delta _{\mathcal{T}_{{\small 0}}}=\partial \left( \Delta _{{\small CP}^{%
{\small 3}}}\right) $ & , & $\mathcal{T}_{{\small 0}}\sim \partial \left(
CP^{3}\right) _{toric}$ & ,%
\end{tabular}%
\end{equation}%
one can construct the leading terms of the family $\left\{ \mathcal{T}_{%
{\small n}}\right\} $ of the blown ups of the complex tetrahedral surface
just by help of the power of toric geometry. Indeed, using the toric
relation $\Delta _{\mathcal{T}_{{\small 0}}}=\partial \left( \Delta _{%
{\small CP}^{{\small 3}}}\right) $, one sees that the toric action $%
U^{3}\left( 1\right) $ generated by translations on the fiber $T^{3}$ of $%
CP^{3}$ (\ref{fbb}) has fix points associated with shrinking p- cycles in $%
T^{3}$. These are:\newline
(\textbf{1}) the divisors of $CP^{3}$; in particular for the four basic $%
\mathcal{D}_{a}$ given by eqs(\ref{d}). On these basic divisors, a 1-cycle
of the 3-torus $T^{3}$ fibration in the bulk of $\Delta _{{\small CP}^{%
{\small 3}}}$ shrinks to zero. As such one is left with $T^{2}$ fibers on
the $\mathcal{D}_{a}$'s as well as a $U\left( 1\right) \times U\left(
1\right) $ toric action as a residual subsymmetry of the $U^{3}\left(
1\right) $ symmetry of the bulk geometry:%
\begin{equation}
\begin{tabular}{llllll}
$CP^{3}$ &  & : &  & basic divisors $\mathcal{D}_{a}$ &  \\ 
$T^{3}$ &  & $\rightarrow $ &  & $\quad \qquad T^{2}$ &  \\ 
$U^{3}\left( 1\right) $ &  & $\rightarrow $ &  & $\qquad U\left( 1\right)
\times U\left( 1\right) $ & 
\end{tabular}%
\end{equation}%
(\textbf{2}) the edges $\Sigma _{\left( {\small ab}\right) }$ of the
tetrahedron on which 2-cycles of $T^{3}$ shrink to zero. Recall that these
edges, which are described by projective lines, are given by the following
intersections,%
\begin{equation}
\begin{tabular}{llll}
$\Sigma _{\left( {\small ab}\right) }$ & $=$ & $\mathcal{D}_{a}\cap \mathcal{%
D}_{b}$ & .%
\end{tabular}
\label{ed}
\end{equation}%
Being toric submanifolds; the complex codimension one divisors $\mathcal{D}%
_{a}$ have as well a toric fibration which we write as follows:%
\begin{equation}
\begin{tabular}{lll}
$T_{a}^{2}$ & $\rightarrow $ & $D_{a}$ \\ 
&  & $\downarrow \pi _{a}$ \\ 
&  & $\Delta _{{\small D}_{{\small a}}}$%
\end{tabular}
\label{fd}
\end{equation}%
where the toric polytope describing $\Delta _{{\small D}_{{\small a}}}$ is a
triangle. Similarly, the intersecting curve $\Sigma _{\left( {\small ab}%
\right) }$ of the two divisors $D_{a}$ and $D_{b}$ is also toric with the
typical fibration%
\begin{equation}
\begin{tabular}{lll}
$S_{\left( ab\right) }^{1}$ & $\rightarrow $ & $\Sigma _{\left( {\small ab}%
\right) }$ \\ 
&  & $\downarrow \pi _{\left( {\small ab}\right) }$ \\ 
&  & $\Delta _{\Sigma _{\left( {\small ab}\right) }}$%
\end{tabular}
\label{eg}
\end{equation}%
where now $\Delta _{\Sigma _{\left( {\small ab}\right) }}$ is represented by
a segment of a straight line. As such, along the curves $\Sigma _{\left( 
{\small ab}\right) }$ the bulk 3- cycles of $T^{3}$ shrinks down to a 1-
cycle fibers $S_{\left( ab\right) }^{1}$ fibers and the $U^{3}\left(
1\right) $ bulk toric action gets reduced to $U\left( 1\right) $. 
\begin{equation}
\begin{tabular}{llllll}
$\ CP^{3}$ &  &  &  & edges $\Sigma _{\left( {\small ab}\right) }$ &  \\ 
$\ \ T^{3}$ &  & $\rightarrow $ &  & $\qquad S_{\left( {\small ab}\right)
}^{1}$ &  \\ 
$U^{3}\left( 1\right) $ &  & $\rightarrow $ &  & $\qquad U_{\left( {\small ab%
}\right) }\left( 1\right) $ & 
\end{tabular}
\label{46}
\end{equation}%
At each point of these projective lines $\Sigma _{\left( {\small ab}\right)
} $ lives then an ordinary $A_{1}$ type singularity associated with the
shrinking of $T^{2}$ whose blow up is done in terms of a real two sphere.%
\newline
(\textbf{3}) the vertices $P_{\left( abc\right) }$ of the tetrahedron given
by the tri- intersection,%
\begin{equation}
\begin{tabular}{llll}
$P_{\left( {\small abc}\right) }$ & $=$ & $\mathcal{D}_{a}\cap \mathcal{D}%
_{b}\cap \mathcal{D}_{c}$ & .%
\end{tabular}
\label{int}
\end{equation}%
At these four vertices, the 3-cycle $T^{3}$ in the bulk geometry shrinks
completely to zero and one is left with a larger singularity involving three
intersecting ordinary $A_{1}$ type singularities which might be thought of
as the affine $A_{2}$ type singularity depicted in the figure (\ref{A2}).
Using the toric fibrations (\ref{fd}) of the basic divisors $\mathcal{D}_{a}$%
, we clearly see that each ordinary A$_{1}$ singularity is associated with
the shrinking of the $T_{a}^{2}$ torus at the tri- vertex intersection (\ref%
{int}).

\ \ \newline
With these features on the toric projective space CP$^{3}$ and their links
to the toric tetrahedral surface $\mathcal{T}_{{\small 0}}$ in mind, we turn
now to study the toric blow ups of the tetrahedron.

\subsubsection{Blow ups of CP$^{\emph{3}}$}

By mimicking the analysis of section 2 regarding the construction of the 
\emph{eight} del Pezzo surfaces $dP_{n}$ from the projective plane $CP^{2}$
and using group theoretical arguments\textrm{\footnote{%
Notice that $CP^{2}\subset C^{3}$ with dimension of the structure group as $%
\dim $ $SU\left( 3\right) =8$. We also have $CP^{3}\subset C^{4}$ with $\dim 
$ $SU\left( 4\right) =15$.}}, one learns that we may a priori perform up to
fifteen blow ups of points in $CP^{3}$ by projective planes. In these blow
ups, the fifteen points which we denote as 
\begin{equation}
\begin{tabular}{llll}
$P_{1},...,P_{15}$ & $\in $ & $CP^{3}$ & ,%
\end{tabular}%
\end{equation}%
get replaced by exceptional projective planes $\digamma _{i}$; $i=1,...,15$.
Because the complex dimension of $CP^{3}$ is odd, we don't have a self dual
homological mid- class and so the derivation of the number \emph{15} need a
little bit more work than in the $CP^{2}$ case. A way to get this number is
to compute the pairing product $\left\langle \Omega _{k}\Omega _{k}^{\ast
}\right\rangle $ of the following real 4- cycle $\Omega _{k}$ and its dual
2- cycle $\Omega _{k}^{\ast }$, 
\begin{equation}
\begin{tabular}{llll}
$\Omega _{k}$ & $=$ & $4G-\sum_{i=1}^{k}\digamma _{i}$ & , \\ 
$\Omega _{k}^{\ast }$ & $=$ & $4H-\sum_{i=1}^{k}E_{i}$ & .%
\end{tabular}%
\end{equation}%
In these relations $G$ is a hyperplane in $CP^{3}$ and $\digamma _{i}$ are
the generators of the blow ups. The generators $H\equiv G^{\ast }$ and $%
E_{i}\equiv \digamma _{i}^{\ast }$ are respectively the dual classes of $G$
and $\digamma _{i}$ satisfying the following pairing products%
\begin{equation}
\begin{tabular}{lllllll}
$\left\langle H,G\right\rangle $ & $=1$ & , &  & $\left\langle
E_{i},\digamma _{j}\right\rangle $ & $=-\delta _{ij}$ & , \\ 
$\left\langle G,G\right\rangle $ & $=0$ & , &  & $\left\langle
E_{i},F_{j}\right\rangle $ & $=0$ & , \\ 
$\left\langle H,H\right\rangle $ & $=0$ & , &  & $\left\langle \digamma
_{i},\digamma _{j}\right\rangle $ & $=0$ & .%
\end{tabular}%
\end{equation}%
Using these relations, we can compute the product $\left\langle \Omega
_{k}\Omega _{k}^{\ast }\right\rangle $ in terms of the positive integer $k$.
We find 
\begin{equation}
\left\langle \Omega _{k}\Omega _{k}^{\ast }\right\rangle =16-k.
\end{equation}%
Positivity of this pairing product requires that the integer $k$ should be
less than $16$. From this result, we learn that the complex tetrahedral
surface $\mathcal{T}_{0}$ \ has a family $\left\{ \mathcal{T}_{k}\right\} $
of \emph{fifteen} cousins%
\begin{equation}
\begin{tabular}{llllll}
$\mathcal{T}_{1}$ & , & \ldots & , & $\mathcal{T}_{15}$ & ,%
\end{tabular}%
\end{equation}%
obtained by blown ups of isolated points of $\mathcal{T}_{0}$\ \ by
projective planes $CP^{2}$. We will see later that the complex tetrahedral
surface $\mathcal{T}_{0}$ \ has a second family $\left\{ \mathcal{T}%
_{m}^{\prime }\right\} $ of \emph{thirty five} cousins. But before coming to
that notice the complex codimension one divisor $\mathcal{T}_{{\small 0}}$
of the complex projective space $CP^{3}$ is described by the real 4- cycles%
\begin{equation}
\Omega _{0}=4G,  \label{4g}
\end{equation}%
where, in toric language, the number \emph{4} in above relation refers to
the four basic divisors $\mathcal{D}_{a}$. Similarly, we have the dual class 
$\Omega _{0}^{\ast }=4H$ associated with the classes of complex lines normal
to the class of the complex surfaces $\mathcal{D}_{a}$ of the complex three
dimension space $CP^{3}$.\newline
Regarding the second family $\left\{ \mathcal{T}_{m}^{\prime }\right\} $ of
cousins of the complex tetrahedral surface $\mathcal{T}_{0}$, notice that
along with the divisor class $\Omega _{0}$ given by eq(\ref{4g}), we may
also define the 2- cycle class $\Upsilon _{0}$ associated with the six edges 
$\Sigma _{\left( ab\right) }$ of the tetrahedron,%
\begin{equation}
\Upsilon _{0}=6H.
\end{equation}%
Its dual class is given by the real 4- cycle $\Upsilon _{0}^{\ast }=6G$ and
it describes the class of the six complex surfaces $\Gamma _{\left(
ab\right) }$ in $CP^{3}$ that are normal to the edges $\Sigma _{\left(
ab\right) }$. Moreover, using the exceptional curves $E_{i}$, one may define
in general the following real 2- cycles class 
\begin{equation}
\Upsilon _{n}=6H-\sum_{i=1}^{n}E_{i},
\end{equation}%
where a priori $n$ is a positive integer. Computing the pairing $%
\left\langle \Upsilon _{n}\Upsilon _{n}^{\ast }\right\rangle $ where $%
\Upsilon _{n}^{\ast }$\ stands for the dual 4-cycle class which reads in
terms of the generators $G$ and $\digamma _{i}$\ like $\Upsilon _{n}^{\ast
}=6G-\sum_{i=1}^{n}\digamma _{i}$, we get 
\begin{equation}
\left\langle \Upsilon _{n}\Upsilon _{n}^{\ast }\right\rangle =36-n,
\end{equation}%
whose positivity require that $n$ should be less than $36$. From this
result, we learn that we may perform up to \emph{35} blow ups by projective
line in $CP^{3}$; these are precisely the second family of cousins of the
complex tetrahedron 
\begin{equation}
\begin{tabular}{llllll}
$\mathcal{T}_{1}^{\prime }$ & , & \ldots & , & $\mathcal{T}_{35}^{\prime }$
& .%
\end{tabular}%
\end{equation}%
Furthermore, we may define as well generic real 4- cycles $\left[ C_{a}%
\right] $ in the complex three dimension space $CP^{3}$. They are given by
the following linear combination%
\begin{equation}
\begin{tabular}{llll}
$C_{a}$ & $=$ & $n_{a}G-\sum_{i}m_{ai}\digamma _{i}$ & ,%
\end{tabular}%
\end{equation}%
with%
\begin{equation}
\begin{tabular}{llll}
$\int_{G}\omega ^{2}=+1$ & , & $\int_{\digamma _{i}}\omega ^{2}=-1$ & ,%
\end{tabular}%
\end{equation}%
and where the 2- form $\omega $\ is as in eqs(\ref{39}-\ref{40}) and where $%
n_{a}$ and $m_{ai}$ are integers. By duality, we also have the real 2-
cycles basis $\left[ \Sigma _{a}\right] \equiv \left[ C_{a}^{\ast }\right] $
in $CP^{3}$ which are given by the following linear combination%
\begin{equation}
\begin{tabular}{llll}
$\Sigma _{a}$ & $=$ & $n_{a}H-\sum_{i}m_{ai}E_{i}$ & ,%
\end{tabular}%
\end{equation}%
with%
\begin{equation*}
\begin{tabular}{llll}
$\int_{H}\omega =+1$ & , & $\int_{E_{i}}\omega =-1$ & .%
\end{tabular}%
\end{equation*}%
The real 2-cycles $\Sigma _{a}$ are in some sense the normal to the real
4-cycles $C_{a}$ in the complex space $CP^{3}$. Their intersection is given
by the pairing product; we have:%
\begin{equation}
\left\langle C_{a}\Sigma _{b}\right\rangle =n_{a}n_{b}-\sum_{i}m_{ai}m_{bi}.
\end{equation}%
We end these comments by recalling that alike in the blowing up of $CP^{2}$,
here also we have the two kinds of blow ups: toric and non toric. In what
follows, we consider the toric blow ups concerning the blowing up of the
vertices and the edges of the tetrahedron. The blowing of the vertices will
be done in terms of projective planes while those of the edges will be done
in terms of projective line.

\subsubsection{Blowing up the vertices}

\begin{figure}[tbph]
\begin{center}
\hspace{0cm} \includegraphics[width=5cm]{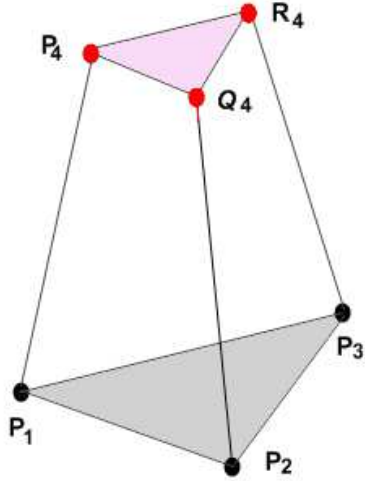}
\end{center}
\par
\vspace{-0.5 cm}
\caption{Surface $\mathcal{T}_{\emph{1}};$ it is given by the 
{\protect\small blow up of the vertex P}$_{{\protect\small 4}}$%
{\protect\small \ of the tetrahedron by a projective plane }${\protect\small %
CP}^{{\protect\small 2}}${\protect\small \ which is described by the
triangle }$\left[ P_{4}Q_{4}R_{4}\right] ${\protect\small .} }
\label{BV1}
\end{figure}

\emph{One blow up: geometry }$\mathcal{T}_{\emph{1}}$\newline
The blow up of one of the four vertices of the tetrahedron $\Delta _{%
\mathcal{T}_{{\small 0}}}$; say the fourth vertex $P_{4}$ in the diagram (%
\ref{BV}); is depicted in figure (\ref{BV1}).

\ \ \newline
In toric language, the blow up of the point $P_{4}$ by a projective plane
amounts to replace the vertex $P_{4}$ of the tetrahedron by a triangle $%
\left[ P_{4}Q_{4}R_{4}\right] $, that is making the substitution%
\begin{equation}
P_{4}\qquad \rightarrow \qquad \left[ P_{4}Q_{4}R_{4}\right] .  \label{bu}
\end{equation}%
As a consequence the tetrahedron $\Delta _{\mathcal{T}_{{\small 0}}}$ gets
deformed to a complex geometry with toric graph $\Delta _{\mathcal{T}_{%
{\small 1}}}$ having five faces. These faces are as follows: \newline
(\textbf{i}) two triangles%
\begin{equation}
\begin{tabular}{llll}
$\left[ P_{1}P_{2}P_{3}\right] $ & $,$ & $\left[ P_{4}Q_{4}R_{4}\right] $ & ,%
\end{tabular}%
\end{equation}%
representing two non intersecting projective planes $CP_{1}^{2}$ and $%
CP_{2}^{2}$. This non intersecting property of the two projective planes is
easily read in the toric geometry language by determining the intersection
of the above triangles:%
\begin{equation}
\begin{tabular}{llll}
$\left[ {\small P}_{1}{\small P}_{2}{\small P}_{3}\right] \cap \left[ 
{\small P}_{4}{\small Q}_{4}{\small R}_{4}\right] $ & $=$ & $\emptyset $ & .%
\end{tabular}%
\end{equation}%
(\textbf{ii}) three quadrilaterals%
\begin{equation}
\begin{tabular}{llllll}
$\left[ P_{1}P_{2}P_{4}Q_{4}\right] $ & , & $\left[ P_{1}P_{3}P_{4}R_{4}%
\right] $ & , & $\left[ P_{2}P_{3}Q_{4}R_{4}\right] $ & ,%
\end{tabular}%
\end{equation}%
describing three intersecting del Pezzo surfaces $dP_{1}^{\left( 1\right) }$%
, $dP_{1}^{\left( {\small 2}\right) }$ and $dP_{1}^{\left( {\small 3}\right)
}$. These intersections may be directly read from the polytope $\Delta _{%
\mathcal{T}_{{\small 1}}}$ of the figure (\ref{BV1}). We have:%
\begin{equation}
\begin{tabular}{llll}
$\left[ P_{1}P_{4}\right] $ & $=$ & $\left[ {\small P}_{1}{\small P}_{2}%
{\small P}_{4}{\small Q}_{4}\right] \cap \left[ {\small P}_{1}{\small P}_{3}%
{\small P}_{4}{\small R}_{4}\right] $ & , \\ 
$\left[ P_{2}Q_{4}\right] $ & $=$ & $\left[ {\small P}_{1}{\small P}_{2}%
{\small P}_{4}{\small Q}_{4}\right] \cap \left[ {\small P}_{2}{\small P}_{3}%
{\small Q}_{4}{\small R}_{4}\right] $ & , \\ 
$\left[ P_{3}R_{4}\right] $ & $=$ & $\left[ {\small P}_{1}{\small P}_{3}%
{\small P}_{4}{\small R}_{4}\right] \cap \left[ {\small P}_{2}{\small P}_{3}%
{\small Q}_{4}{\small R}_{4}\right] $ & ,%
\end{tabular}%
\end{equation}%
Notice that using the generator $G$ and $\digamma _{i}$, we can define the
blow up surface represented by the polytope $\Delta _{\mathcal{T}_{{\small 1}%
}}$ in terms of the "canonical 4- cycle" as follows: 
\begin{equation}
\Omega _{1}=4G-\digamma _{1},
\end{equation}%
where $\digamma _{1}$ generates the blow up (\ref{bu}). Notice also the
emergence of the del Pezzo surfaces $dP_{1}$ into the geometry of the blown
up of the complex tetrahedral surface. This result is not a strange thing
since it was expected from the analysis of section 2 since after all the
blown up of the tetrahedron involves implicitly the blowing up of projective
planes constituting the tetrahedral surface.\ 

\begin{figure}[tbph]
\begin{center}
\hspace{0cm} \includegraphics[width=5cm]{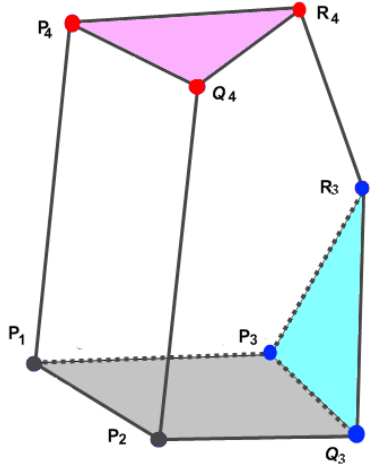}
\end{center}
\par
\vspace{-0.5 cm}
\caption{Surface $\mathcal{T}_{\emph{2}}$; it is given by the blow up of the
vertices $P_{3}$ and $P_{{\protect\small 4}}$\ of the tetrahedron by two
respective projective planes $CP_{1}^{{\protect\small 2}}$\ and $CP_{2}^{%
{\protect\small 2}}$\ described by the triangles $\left[ P_{3}Q_{3}R_{3}%
\right] $ and $\left[ P_{4}Q_{4}R_{4}\right] $. }
\label{BV2}
\end{figure}

\emph{Two blow ups: geometry }$\mathcal{T}_{\emph{2}}$\newline
In the case of the blown up of two vertices of the tetrahedron, say the
third vertex $P_{3}$ and the fourth $P_{4}$ one, we get a geometry $\mathcal{%
T}_{\emph{2}}$ that involves more intersecting del Pezzo surfaces. The toric
graph $\Delta _{\mathcal{T}_{\emph{2}}}$ of this blown up surface is
depicted in figure (\ref{BV2}),

\ \ \newline
The corresponding polytope $\Delta _{\mathcal{T}_{\emph{2}}}$ has six
intersecting faces as reported in the following table, 
\begin{equation}
\begin{tabular}{llllll}
triangles & : & $\left[ {\small P}_{3}{\small Q}_{3}{\small R}_{3}\right] $
& , & $\left[ {\small P}_{4}{\small Q}_{4}{\small R}_{4}\right] $ & , \\ 
quadrilaterals & : & $\left[ {\small P}_{1}{\small P}_{2}{\small P}_{4}%
{\small Q}_{4}\right] $ & , & $\left[ {\small P}_{1}{\small P}_{2}{\small P}%
_{3}{\small Q}_{3}\right] $ & , \\ 
pentagons & : & $\left[ {\small P}_{1}{\small P}_{3}{\small R}_{3}{\small R}%
_{4}{\small P}_{4}\right] $ & , & $\left[ {\small P}_{2}{\small Q}_{3}%
{\small R}_{3}{\small Q}_{4}{\small R}_{4}\right] $ & , \\ 
&  &  &  &  & 
\end{tabular}%
\end{equation}%
from which one may read directly the intersections. The triangles describe
respectively two projective planes $dP_{0}^{\left( {\small 1}\right) }$ and $%
dP_{0}^{\left( {\small 2}\right) }$, the quadrilaterals describe two del
Pezzo $dP_{1}$ surfaces defining the third and the fourth faces $%
dP_{1}^{\left( {\small 3}\right) }$ and $dP_{1}^{\left( {\small 4}\right) }$
and the pentagons are associated with two del Pezzo $dP_{2}$ geometries
giving the fifth and the sixth faces $dP_{2}^{\left( {\small 5}\right) }$
and $dP_{2}^{\left( {\small 6}\right) }$. \newline
Using the real 4- cycle generators $G$ and $\digamma _{i}$, the real 4-
cycle class $\left[ \Omega _{2}\right] $ describing the two blow ups of the
tetrahedron is given by%
\begin{equation}
\Omega _{2}=4G-\digamma _{1}-\digamma _{2}\text{ },
\end{equation}%
where $\digamma _{1}$ and $\digamma _{2}$ generate the blown ups of the
points 
\begin{equation}
\begin{tabular}{llll}
${\small P}_{3}$ & $\rightarrow $ & $\left[ {\small P}_{3}{\small Q}_{3}%
{\small R}_{3}\right] $ & , \\ 
${\small P}_{4}$ & $\rightarrow $ & $\left[ {\small P}_{4}{\small Q}_{4}%
{\small R}_{4}\right] $ & .%
\end{tabular}%
\end{equation}

\begin{figure}[tbph]
\begin{center}
\hspace{0cm} \includegraphics[width=5cm]{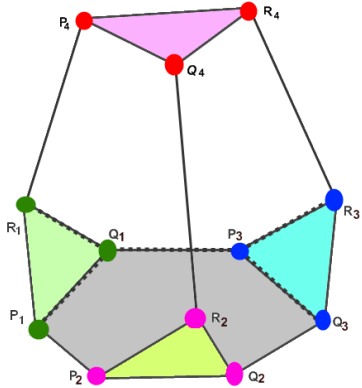}
\end{center}
\par
\vspace{-0.5 cm}
\caption{Geometry $\mathcal{T}_{\emph{4}}$; its given by the blow up of the
four vertices of the tetrahedron by a projective planes $CP^{{\protect\small %
2}}$. }
\label{BV4}
\end{figure}

$\mathcal{T}_{\emph{3}}$ and $\mathcal{T}_{\emph{4}}$ geometries\newline
Similar analysis may be done for the blown up of three and four vertices.
For the blown up of three vertices; say $P_{2}$, $P_{3}$ and $P_{4}$, one
gets a polytope $\Delta _{\mathcal{T}_{\emph{3}}}$ with seven intersecting
faces: (i) three triangles, (ii) three pentagons and (iii) an hexagon,%
\begin{equation}
\begin{tabular}{llllllll}
&  &  &  &  &  &  &  \\ 
triangles & : & $\left[ P_{2}Q_{2}R_{2}\right] $ & , & $\left[ {\small P}_{3}%
{\small Q}_{3}{\small R}_{3}\right] $ & , & $\left[ {\small P}_{4}{\small Q}%
_{4}{\small R}_{4}\right] $ & , \\ 
pentagons & : & $\left[ {\small P}_{1}{\small P}_{2}{\small R}_{2}{\small Q}%
_{4}{\small P}_{4}\right] $ & , & $\left[ {\small P}_{1}{\small P}_{3}%
{\small R}_{3}{\small R}_{4}{\small P}_{4}\right] $ & , & $\left[ {\small P}%
_{2}{\small Q}_{3}{\small R}_{3}{\small Q}_{4}{\small R}_{4}\right] $ & , \\ 
hexagon & : & $\left[ {\small R}_{2}{\small Q}_{2}{\small Q}_{3}{\small R}%
_{3}{\small Q}_{4}{\small R}_{4}\right] $ & . &  &  &  &  \\ 
&  &  &  &  &  &  & 
\end{tabular}%
\end{equation}%
For the blown up of the four vertices of the tetrahedron, one obtains the
geometry depicted in the figure (\ref{BV4}),

\ \ \newline
The resulting toric graph $\Delta _{\mathcal{T}_{\emph{4}}}$ has twelve
vertices and eight faces given by four triangles describing four exceptional
projective planes and four hexagons associated with the del Pezzo surfaces $%
dP_{3}$.

\subsection{Blow ups by CP$^{1}$s}

Here also we restrict our analysis to the toric blow ups concerning the $%
A_{1}$ singularities degenerating along the edges $\Sigma _{\left( ab\right)
}$ eqs(\ref{ed}-\ref{eg}) of the tetrahedron $\Delta _{\mathcal{T}_{{\small 0%
}}}$. To that purpose, notice first that contrary to the familiar cases
where 2- cycle degeneracies takes place at isolated points on manifolds,
here the $A_{1}$ singularity take place along the edges of the tetrahedron $%
\Delta _{\mathcal{T}_{{\small 0}}}$; that is for all those points $P$ of the
tetrahedron $\Delta _{\mathcal{T}_{{\small 0}}}$ where a 2- torus shrinks
down to zero.

\subsubsection{From a edge to a $dP_{1}$ surface}

First recall that in the case of singularities at isolated points the blow
up is achieved in terms a complex surface namely a projective plane $CP^{2}$%
. Here we complete this study by showing that for the case of the A$_{1}$
singularity on the edges, the blow up is achieved as well in terms of a
complex surface but this time in terms of $dP_{1}$ surface.\newline
Indeed, given a segment $\left[ AB\right] $ describing a projective line $%
CP^{1}$ where at each point $P\in \left[ AB\right] $ lives a $A_{1}$
singularity, the blow up of such singularity consists to replace each point $%
P$ by a segment $\left[ PQ\right] $ as it is usually done,%
\begin{equation}
\begin{tabular}{llll}
$P$ & $\rightarrow $ & $\left[ PQ\right] $ & .%
\end{tabular}%
\end{equation}%
This means that each singular point $P$ is substituted by a rational curve.
Doing so for all points $P$ belonging to the segment $\left[ AB\right] $, we
end with the quadrilateral%
\begin{equation}
\left[ AB\right] \times \left[ CD\right] .
\end{equation}%
The blowing up of an edge $\left[ P_{a}P_{b}\right] $ of the tetrahedron $%
\Delta _{\mathcal{T}_{{\small 0}}}$ of the figure (\ref{BV}) corresponds in
the language of toric graphs to the replacement 
\begin{equation}
\begin{tabular}{llll}
$\left[ P_{a}P_{b}\right] $ & $\rightarrow $ & $P_{a}P_{b}\times \left[
Q_{a}Q_{b}\right] \sim \left[ P_{a}Q_{a}P_{b}Q_{b}\right] $ & .%
\end{tabular}%
\end{equation}%
In complex geometry, the blow up of an edge $\Sigma _{\left( ab\right) }$ $%
\sim $ $CP^{1}$ of the complex tetrahedral surface amounts to replace the
complex projective line $CP^{1}$ by a del Pezzo surface $dP_{1}$:%
\begin{equation}
\begin{tabular}{llll}
$CP^{1}$ & $\rightarrow $ & $dP_{1}$ & .%
\end{tabular}%
\end{equation}%
Let us apply this construction to the blowing up of two independent edges of
the tetrahedron $\Delta _{\mathcal{T}_{{\small 0}}}$; say $\left[ P_{1}P_{3}%
\right] $ and $\left[ P_{2}P_{4}\right] $ with $\left[ P_{1}P_{3}\right]
\cap \left[ P_{2}P_{4}\right] =\emptyset $. \newline
First, we study the blow up of the edge $\left[ P_{2}P_{4}\right] \in \Delta
_{\mathcal{T}_{{\small 0}}}$ of the figure (\ref{BV}) and then we consider
the blow up of the two edges $\left[ P_{1}P_{3}\right] $ and $\left[
P_{2}P_{4}\right] $.

\subsubsection{Blowing up the edge $\left[ P_{2}P_{4}\right] $}

\begin{figure}[tbph]
\begin{center}
\hspace{0cm} \includegraphics[width=5cm]{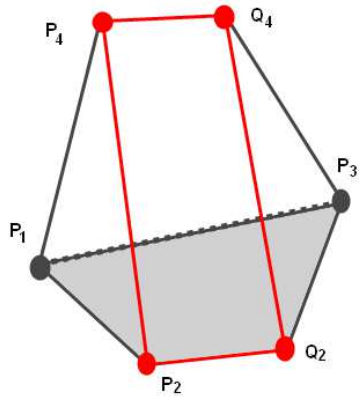}
\end{center}
\par
\vspace{-0.5 cm}
\caption{Surface $\mathcal{T}_{1}^{\prime }$; it is given by the blow up of
a edge $\left[ P_{2}P_{4}\right] $\ of the graph fig(\protect\ref{BV}) by a
projective planes $CP^{{\protect\small 1}}$. The resulting geometry is a del
Pezzo surface described by the polygon $\left[ P_{2}Q_{2}P_{4}Q_{4}\right] $%
. The full geometry has five faces del Pezzo surfaces whose intersections
are directly read fom the toric graph.}
\label{SE1}
\end{figure}

The blow up of the edge $\left[ P_{2}P_{4}\right] $ of the tetrahedron of
the graph (\ref{BV}) is depicted in the figure (\ref{SE1}). The edge $\left[
P_{2}P_{4}\right] $, which represents a complex projective line, has been
replaced by the quadrilateral $\left[ P_{2}Q_{2}P_{4}Q_{4}\right] $:%
\begin{equation}
\begin{tabular}{llll}
$\left[ P_{2}P_{4}\right] $ & $\rightarrow $ & $\left[ P_{2}Q_{2}P_{4}Q_{4}%
\right] $ & .%
\end{tabular}%
\end{equation}%
The obtained polytope has five faces and six vertices where meet three faces
as well as three edges. Regarding the five faces, we have: \newline
\textbf{(i) }two triangles $\left[ P_{1}P_{2}P_{4}\right] $ and $\left[
Q_{2}P_{3}Q_{4}\right] $ describing two projective planes.\newline
\textbf{(ii) }three quadrilaterals $\left[ P_{1}P_{2}Q_{2}P_{3}\right] $, $%
\left[ P_{2}Q_{2}P_{4}Q_{4}\right] $, and $\left[ P_{1}P_{3}P_{4}Q_{4}\right]
$ describing $dP_{1}$ surfaces. These del Pezzo surfaces intersects mutually
and intersect as well with the projective planes.

\subsubsection{Blowing up the edge $\left[ P_{1}P_{3}\right] $ and $\left[
P_{2}P_{4}\right] $}

\begin{figure}[tbph]
\begin{center}
\hspace{0cm} \includegraphics[width=6cm]{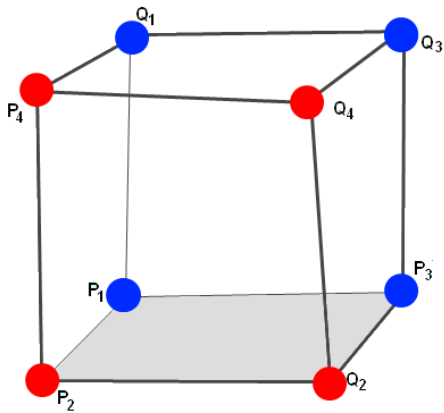}
\end{center}
\par
\vspace{-0.5 cm}
\caption{Surface $\mathcal{T}_{2}^{\prime }$: it is given by the blow up of
two edges of the tetrahedron fig(\protect\ref{BV}): $\left[ P_{1}P_{3}\right]
\rightarrow \left[ P_{1}Q_{1}P_{3}Q_{3}\right] $ and $\left[ P_{2}P_{4}%
\right] \rightarrow \left[ P_{2}Q_{2}P_{4}Q_{4}\right] $. The resulting
geometry has eight vertices, twelve edges and six faces describing six
intersecting del Pezzo surfaces $dP_{{\protect\small 1}}$s. The
intersections are directly read on this graph.}
\label{SE21}
\end{figure}

The blow up of two edges of the tetrahedron by projective lines is depicted
in the figure (\ref{SE21}). The edges $\left[ P_{1}P_{3}\right] $ and $\left[
P_{2}P_{4}\right] $ have been replaced by the quadrilaterals $\left[
P_{1}Q_{1}P_{3}Q_{3}\right] $ and $\left[ P_{2}Q_{2}P_{4}Q_{4}\right] $. 
\newline
The obtained polytope has six quadrilateral faces describing six
intersecting del Pezzo surfaces $dP_{1}$.

\section{Conclusion and discussions}

\begin{figure}[tbph]
\begin{center}
\hspace{0cm} \includegraphics[width=14cm]{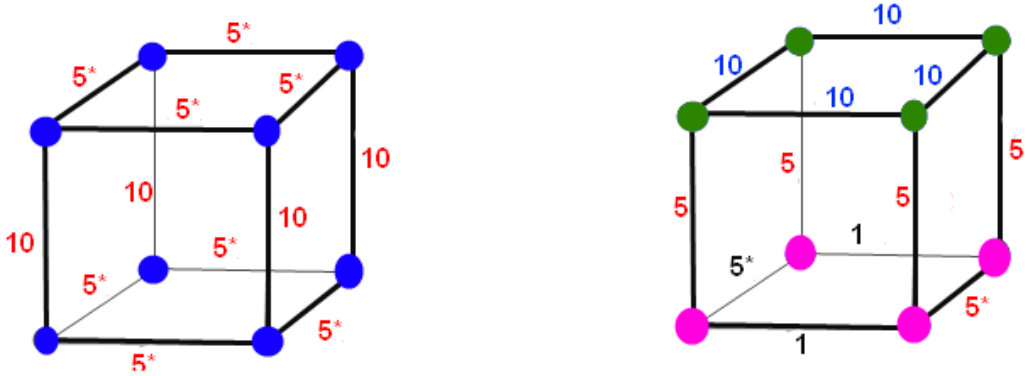}
\end{center}
\par
\vspace{-0.5 cm}
\caption{{\protect\small On left we have the quiver gauge diagram of an SU}$%
\left( 5\right) ${\protect\small \ GUT-like model with eight Yukawa
couplings type }${\protect\small 5}_{a}^{\ast }{\protect\small \times 5}%
_{b}^{\ast }{\protect\small \times 10}_{c}${\protect\small . On the right
the quiver graph of an SU}$\left( 5\right) ${\protect\small \ GUT-like model
with four Yukawa couplings type }${\protect\small 5}_{a}^{\ast }%
{\protect\small \times 5}_{b}^{\ast }{\protect\small \times 10}_{c}$%
{\protect\small \ and four others of type }$5_{a}^{\ast }\times 5_{b}\times
1_{c}${\protect\small . These two }$SU\left( 5\right) ${\protect\small \
GUT-like models have exotic fields.}}
\label{SE}
\end{figure}

Motivated by F-theory- GUT models building along the line of the BHV
approach \textrm{\cite{B1,B2,B3}} and guided by special properties of the
toric fibration of complex surfaces, we have studied in this paper two
families of blowing up of the complex tetrahedral surfaces $\mathcal{T}_{%
{\small 0}}$. These families, which were respectively denoted as $\mathcal{T}%
_{{\small n}}$ with $n\leq 15$ and $\mathcal{T}_{{\small k}}^{\prime }$ with 
$k\leq 35$ are as follows:\newline
\textbf{1}) the blowing up of the complex three dimension space $CP^{3}$ up
to fifteen isolated points by projective planes $CP^{2}$. Four of these blow
ups are of toric type and have been explicitly studied by using the power of
the standard toric graph representation and n-simplex description. If
denoting by 
\begin{equation}
\begin{tabular}{llll}
$\left( CP^{3}\right) _{n,0}$ & $,$ & $n=1,...,15$ & $,$%
\end{tabular}%
\end{equation}
the blowing ups of the $CP^{3}$ at n isolated points, then the link between
these $\left( CP^{3}\right) _{n,0}$s and the blown up tetrahedral surfaces $%
\mathcal{T}_{{\small n}}$ is given by means of toric geometry where roughly
the $\mathcal{T}_{{\small n}}$s appear as their toric boundary; see also
footnote (2).\newline
Notice that viewed from the $CP^{3}$ side, the toric singularity at the
tetrahedron vertices $P_{\left( abc\right) }$ is given the shrinking of a
real 3-torus of the fibration $CP^{3}\sim T^{3}\times \Delta _{{\small CP}^{%
{\small 3}}}$. On the complex tetrahedral surface side however, the visible
toric singularity at%
\begin{equation}
P_{\left( {\small abc}\right) }=\mathcal{D}_{a}\cap \mathcal{D}_{b}\cap 
\mathcal{D}_{c},  \label{C1}
\end{equation}%
is given by simultaneous shrinking of three 2- tori namely the\ shrinking of
the toric fibers $T_{a}^{2},$ $T_{b}^{2}$ and $T_{c}^{2}$ of the respective
divisors $\mathcal{D}_{a},$ $\mathcal{D}_{b}$ and $\mathcal{D}_{c}$; see eq(%
\ref{fd}). \newline
\textbf{2}) the blowing up of $CP^{3}$ up to thirty five projective lines.
Six of these blow ups are of toric type. These blow ups are different from
the $\left( CP^{3}\right) _{n,0}$ ones since they concern the blown up of A$%
_{1}$ singularities. We may refer to them as,%
\begin{equation}
\begin{tabular}{llll}
$\left( CP^{3}\right) _{0,k}$ & $,$ & $k=1,...,35$ & $.$%
\end{tabular}%
\end{equation}
in order to distinguish them of the previous $\left( CP^{3}\right) _{n,0}$
family. In this case, the toric singularity living on the tetrahedron edges 
\begin{equation}
\Sigma _{\left( ab\right) }=\mathcal{D}_{a}\cap \mathcal{D}_{b}  \label{C2}
\end{equation}%
is associated with the shrinking of a real 2-torus of the fibration $%
CP^{3}\sim T^{3}\times \Delta _{{\small CP}^{{\small 3}}}$ down to $S^{1}$
as shown on eq(\ref{46}). Viewed from the divisors $\mathcal{D}_{a}$ and $%
\mathcal{D}_{b}$, the singularity on the edge corresponds to the shrinking
of a 1-cycle along the intersection of $\mathcal{D}_{a}$ and $\mathcal{D}%
_{b} $. \newline
Through this study we learned a set of special features amongst which the
two following: \newline
\textbf{a}) the toric blown ups $\mathcal{T}_{{\small n}}$ and $\mathcal{T}_{%
{\small k}}^{\prime }$ of the complex tetrahedral surface $\mathcal{T}_{%
{\small 0}}$ are mainly given by intersecting del Pezzo surfaces $dP_{k}$.
This property is expected from general arguments since the blowing of the
tetrahedron 
\begin{equation}
\mathcal{T}=\cup _{a}\mathcal{D}_{a}
\end{equation}%
together with the relations (\ref{C1}-\ref{C2}), amounts to blowing the
divisors $\mathcal{D}_{a}$. But these divisors homeomorphic to $CP^{2}$s
embedded in $CP^{3}$. We have checked this property for the toric blow ups
type; but \ we don't have yet the answer whether this result is true as well
for the non toric blow ups. \newline
\textbf{b}) Toric geometry has a nice feature which can be used in the
engineering of F-theory GUT- like models building. In going from the faces
to the vertices of the tetrahedron, cycles of the toric fibers shrink down
as shown in the following table

\begin{equation}
\begin{tabular}{|lllllll|}
\hline
&  &  &  &  &  &  \\ 
tetrahedron $\Delta _{\mathcal{T}_{n}}$ & : & \ \ \ \ \ \ \ \ faces \  & \ \
\ \ \  & \ \ edges &  & vertices \\ 
toric fibers & : & $\ \ \ \ \ \ \ \ \ \ \ T^{2}$ & $\ \rightarrow $ & $\ \ \
\ \ S^{1}$ & $\rightarrow $ & \ \ 0 \\ 
toric symmetries & : & $\ \ \ \ U\left( 1\right) \times U\left( 1\right) $ \
\ \ \ \  & $\ \rightarrow $ & $\ \ U\left( 1\right) $\ \ \  & $\rightarrow $
& \ \ 0 \\ 
&  &  &  &  &  &  \\ \hline
\end{tabular}%
\end{equation}

\ \ \newline
In the field theory language, these shrinking generate massless modes which
may be interpreted in terms of massless gauge fields and so gauge symmetry
enhancements at the level of the 4D space time effective field theory. More
precisely, given a gauge symmetry $G_{r}$ that is visible 4D space time, the
gauge symmetry associated with the faces $\mathcal{D}_{a}$ of the
tetrahedron and its blow ups would be%
\begin{equation}
G_{r}\times U\left( 1\right) \times U\left( 1\right) ,  \label{sy}
\end{equation}%
where the $U\left( 1\right) $ factors may be interpreted in terms of branes
wrapping cycles in the toric fibration. The bulk invariance (\ref{sy}) gets
enhanced to a $G_{r+1}\times U\left( 1\right) $ invariance on the edges $%
\Sigma _{\left( ab\right) }$ and further to a $G_{r+2}$ gauge symmetry at
the vertices $P_{\left( abc\right) }$. \newline
In the case where $G_{r}=SU\left( 5\right) $ for example, the gauge enhanced
symmetry on the edges could be either $SU\left( 6\right) $ or $SO\left(
10\right) $ and at the vertices it may be one of the following enhanced
gauge symmetries%
\begin{equation}
\begin{tabular}{llllll}
$SU\left( 7\right) $ & , & $SO\left( 12\right) $ & , & $E_{6}$ & .%
\end{tabular}%
\end{equation}%
We end this conclusion by adding a comment regarding the way the tetrahedron
surface and its blown up cousins $\mathcal{T}_{n}$ and $\mathcal{T}%
_{k}^{\prime }$\ could be used in practice. They should be thought of as the
base surface of the elliptically K3 fibered Calabi-Yau four- folds in the
F-theory compactification to 4D space time,%
\begin{equation}
\begin{tabular}{lll}
Y & $\rightarrow $ & CY4 \\ 
&  & $\downarrow \pi _{n}$ \\ 
&  & $\mathcal{T}_{{\small n}}$%
\end{tabular}%
\end{equation}%
These complex surfaces are wrapped by seven branes with intersections along
the edges and at the vertices. On the edges localize chiral matters $\Phi
_{R_{a}}^{a}$ in bi-fundamental representations while at the vertices of the
toric graphs live a 4D $\mathcal{N}=1$ supersymmetric Yukawa couplings with
chiral potential,%
\begin{equation}
W_{Yuk}=\int d^{4}xd^{2}\theta \left( \sum_{a<b<c}\frac{\lambda _{abc}}{3}%
\Phi _{R_{a}}^{a}\Phi _{R_{b}}^{b}\Phi _{R_{c}}^{c}\right) .
\end{equation}%
where the complex numbers $\lambda _{abc}$ are Yukawa coupling constants. In
this 4D $\mathcal{N}=1$ superspace relation, $\Phi _{R_{a}}^{a}$ stands for
a family of chiral superfields transforming in representations $R_{a}$ of
the gauge group $G_{r}\times U\left( 1\right) \times U\left( 1\right) $ with
the constraint equation 
\begin{equation}
R_{a}\otimes R_{b}\otimes R_{c}=\mathbf{1}\oplus \left( \dbigoplus_{i}%
\mathrm{f}_{abc}^{i}R_{i}\right) ,
\end{equation}%
where $\mathrm{f}_{abc}^{i}$ are some positive integers capturing the
multiplicity of the representation $R_{i}$. If relaxing the BHV model to
include as well those unrealistic F-theory \emph{GUT-like} models by
allowing exotic fields, the blow ups geometries $\mathcal{T}_{n}$ and $%
\mathcal{T}_{k}^{\prime }$ may be used to engineer effective quiver gauge
theories that are embedded in F-theory on Calabi-Yau 4-folds. In this view,
by using for instance the blown surface of figure (\ref{SE21}) and taking 
\begin{equation}
G_{r}=SU\left( 5\right) ,
\end{equation}
we may engineer various 4D $\mathcal{N}=1$ supersymmetric $SU\left( 5\right) 
$ quiver gauge models like the two ones depicted on the figures (\ref{SE}).
For these examples, chiral matters $\Phi _{R_{a}}^{a}$ localizing on each on
of the twelve edges $\Sigma _{\left( ab\right) }$ transform in on of the
following SU$\left( 5\right) $ representations%
\begin{equation}
\begin{tabular}{llllllllll}
$R_{a}$ & $\equiv $ & $1$ & $,$ & $5$ & $,$ & $5^{\ast }$ & , & $10$ & .%
\end{tabular}%
\end{equation}
These representations $R_{a}$, which carry also charges $\left(
q_{a},q_{a}^{\prime }\right) $ under the $U\left( 1\right) \times U\left(
1\right) $ toric symmetry, describe the usual chiral matter $5_{M}^{\ast }$
and $10_{M}$ as well as its Higgs fields $5_{up}$ and $5_{down}^{\ast }$ of
the $SU\left( 5\right) $ GUT model but also exotic matter.\newline
Yukawa couplings localizing at the eight vertices $P_{\left( abc\right) }$
of the graph (\ref{SE21}) are given by 
\begin{equation}
\begin{tabular}{llll}
$5_{a}^{\ast }\times 5_{b}^{\ast }\times 10_{c}$ & , & $5_{a}^{\ast }\times
5_{b}\times 1_{c}$ & ,%
\end{tabular}%
\end{equation}%
where the singlets $1_{c}$ stand for right neutrinos-like and right leptons-
like. The geometric engineering of such kind of quiver gauge theories will
be extensively developed in \textrm{\cite{DMS}.} 

\begin{acknowledgement}
: I would like to thank Dr L.B Drissi for discussions. This research work is
supported by the program Protars III D12/25.
\end{acknowledgement}

\end{document}